# Separation of Bell States

B. C. Sanctuary
Department of Chemistry,
McGill University
Montreal Quebec
Canada

**Abstract**. The four Bell states can be represented by separable coherent states which are products of individual non-hermitian spin operators. In the absence of interactions, the non-hermitian states are predicted to form a new quantum state of spin magnitude $1/\sqrt{2}$ rather than $1/2$. Properties of these states show that an isolated spin is a resonance state with zero net angular momentum, consistent with a point particle. In addition, the Bell states are shown to take on the identical mathematical form when the two spins are bound (local) or unbound (non-local). The bound Bell states are resonances between four states. When the separate, they do so from only one of its resonance states and their ensemble average defines the unbound Bell states. The bound and unbound Bell states have the same mathematical form due to the persistence of the rotationally invariance of $\boldsymbol{\sigma}^1 \cdot \boldsymbol{\sigma}^2$.

## 1. INTRODUCTION

The four Bell states are separable when a specific non-hermiticity is introduced into a spin ½. Although such states are outside the postulates of quantum theory, in this paper some of the properties and consequences are examined. Part of the motivation is the prediction that an isolated spin has a resonance state of magnitude of $1/\sqrt{2}$, and this gives a basis for explaining experiments where quantum theory so far has failed. These are briefly mentioned in the discussion.

Even though the non-hermiticity does not lead to complex eigenvalues, non-the-less some expectation values would be complex if they were used. This is discussed in the Section 4 where it is argued that the non-hermiticity does not survive to the level that expectation values are viable. The non-hermiticity expresses coherence within a single spin and an expectation value cannot be defined for a single entity. This notion of a spin state for a single particle is developed further by the use of decoherence and ensemble averaging over a statistically large number of such single particles and this results in the usual hermitian statistical state operators with corresponding real expectation values. Several cases are treated: isolated single spins; two particle states, *i.e.* the Bell states; and under the influence of an external magnetic field.

Resonance within single structures plays a pivotal role. For an isolated single spin, internal resonance leads to the prediction that the angular momentum cancels leaving a point-like isotropic particle. In the presence of local interactions, only one of the resonances survives and this gives the usual spin as defined by a Stern-Gerlach experiment.

Similarly the two spins that form a singlet state have four resonances and the anisotropic terms also cancel leaving a rotationally invariant contribution which is the same for all such pairs whether local or separated: $\boldsymbol{\sigma}^1 \cdot \hat{\mathbf{n}} \hat{\mathbf{n}}^* \cdot \boldsymbol{\sigma}^2 \propto \boldsymbol{\sigma}^1 \cdot \boldsymbol{\sigma}^2$. Here $\hat{\mathbf{n}}$ is a unit vector that orients a spin in its unique body coordinate frame in 3D physical space. Coincidence experiments on photons measure the correlated pairs $\boldsymbol{\sigma}^1 \cdot \hat{\mathbf{n}}$ and $\boldsymbol{\sigma}^2 \cdot \hat{\mathbf{n}}^*$ and the accumulated effect is the build up of two particle correlation $\boldsymbol{\sigma}^1 \cdot \boldsymbol{\sigma}^2$. Similar interpretations follow for the other Bell states.

Section 2 the four Bell states are written as a sum of product states for two correlated particles, called bi-particles. Section 3 gives some properties of the non-hermitian states and the new pure resonance hermitian spin state of magnitude $1/\sqrt{2}$ is obtained. The results of Sections 2 and 3 are interpreted in Section 4, and a discussion concludes the paper.

## 2. NON-HERMITIAN STATES

A singlet state is commonly expressed in terms of the entangled Bell state as

$$|\psi_{12}^-\rangle = \frac{1}{\sqrt{2}}\left(|+\rangle_Z^1 |-\rangle_Z^2 - |-\rangle_Z^1 |+\rangle_Z^2\right) \tag{2.1}$$

where the $|\pm\rangle_Z^i$ are the usual eigenstates[1] of the Pauli spin operator $\sigma_Z^i$ with eigenvalues of $\pm 1$. Since the singlet is isotropic, the choice of the quantization axis, $\hat{Z}$ (subscript Z), is arbitrary as long as it coincides for both spins. For this reason the LHS of Eq.(2.1) has no dependence on $\hat{Z}$. The statistical, or density, operator[2,3] for this two state spin is given in matrix form by

$$\rho_{\psi_{12}^-} = |\psi_{12}^-\rangle\langle\psi_{12}^-| = \begin{pmatrix} 0 & 0 & 0 & 0 \\ 0 & 1 & -1 & 0 \\ 0 & -1 & 1 & 0 \\ 0 & 0 & 0 & 0 \end{pmatrix}^{12} \tag{2.2}$$

which displays the quantum interference, or coherence, between the off-diagonal elements $|\pm\rangle_Z^1|\mp\rangle_Z^2$ and $|\mp\rangle_Z^1|\pm\rangle_Z^2$. Such terms are responsible for the entanglement[4]. There is no subscript Z on Eq.(2.2) because its form is invariant to the choice of coordinate system in three dimensional physical space[5].

The singlet state is considered to be entangled because it is impossible to represent it in terms of a product of the usual statistical operators for an ensemble of spins ½. Such pure states as formed for example after the particles pass through a Stern-Gerlach filter[6], and are given in terms of the Pauli spin matrix for the $\hat{Z}$ component as,

$$\rho_{Z+} \equiv |+\rangle_Z^1 \langle+|_Z^1 = \frac{1}{2}(I+\sigma_Z) = \begin{pmatrix} 1 & 0 \\ 0 & 0 \end{pmatrix}_Z \quad \rho_{Z-} \equiv |-\rangle_Z^1 \langle-|_Z^1 = \frac{1}{2}(I-\sigma_Z) = \begin{pmatrix} 0 & 0 \\ 0 & 1 \end{pmatrix}_Z \tag{2.3}$$

The quantization axis is in the direction of the external magnetic field, $\hat{Z}$, hence $\hat{X}, \hat{Y}, \hat{Z}$ represents a laboratory coordinate frame. In contrast, since any axis of quantization is acceptable for the singlet state, a specific one particle, or body, coordinate frame by $\hat{x}, \hat{y}, \hat{z}$, is introduced, and related to the laboratory frame by a rotation. The coherence is defined by,

$$\rho_z(111) \equiv \frac{1}{2}(I+\sigma_z+\sigma_x+i\sigma_y) = \begin{pmatrix} 1 & 1 \\ 0 & 0 \end{pmatrix}_z \tag{2.4}$$

which renders the state non-hermitian. A non-hermitian state cannot be described by a wave function, such as Eq.(2.1), so the density operator form, such as in Eqs.(2.2), is used. Equation (2.4) is, however, not a statistical operator but treated here as the state of a single spin ½. The subscript $z$ indicates any body fixed frame in three dimensional real space.

Then it follows that the state operator for a singlet state, Eq.(2.2) can be expressed as a direct product of eight permutations of the above state[5],

$$\rho_{\Psi_{12}^-} = \frac{1}{8}\begin{pmatrix} \left[\begin{pmatrix} 1 & 1 \\ 0 & 0 \end{pmatrix}_z^1 \otimes \begin{pmatrix} 0 & 0 \\ -1 & 1 \end{pmatrix}_z^2 + \begin{pmatrix} 1 & 0 \\ 1 & 0 \end{pmatrix}_z^1 \otimes \begin{pmatrix} 0 & -1 \\ 0 & 1 \end{pmatrix}_z^2\right] + \\ \left[\begin{pmatrix} 0 & 0 \\ -1 & 1 \end{pmatrix}_z^1 \otimes \begin{pmatrix} 1 & 1 \\ 0 & 0 \end{pmatrix}_z^2 + \begin{pmatrix} 0 & -1 \\ 0 & 1 \end{pmatrix}_z^1 \otimes \begin{pmatrix} 1 & 0 \\ 1 & 0 \end{pmatrix}_z^2\right] + \\ \left[\begin{pmatrix} 1 & 0 \\ -1 & 0 \end{pmatrix}_z^1 \otimes \begin{pmatrix} 0 & 1 \\ 0 & 1 \end{pmatrix}_z^2 + \begin{pmatrix} 1 & -1 \\ 0 & 0 \end{pmatrix}_z^1 \otimes \begin{pmatrix} 0 & 0 \\ 1 & 1 \end{pmatrix}_z^2\right] + \\ \left[\begin{pmatrix} 0 & 1 \\ 0 & 1 \end{pmatrix}_z^1 \otimes \begin{pmatrix} 1 & 0 \\ -1 & 0 \end{pmatrix}_z^2 + \begin{pmatrix} 0 & 0 \\ 1 & 1 \end{pmatrix}_z^1 \otimes \begin{pmatrix} 1 & -1 \\ 0 & 0 \end{pmatrix}_z^2\right] \end{pmatrix} = \frac{1}{2}\begin{pmatrix} 0 & 0 & 0 & 0 \\ 0 & 1 & -1 & 0 \\ 0 & -1 & 1 & 0 \\ 0 & 0 & 0 & 0 \end{pmatrix}^{12} \tag{2.5}$$

The symbol $\otimes$ denotes the direct product of the two $2\times 2$ dimensional operator spaces of each spin, 1 and 2. Equation (2.5) shows that the singlet state is separable into a sum of product states. The individual $2\times 2$ states are all pure and non-hermitian. However the two products shown in each of the four square brackets are hermitian and are identified in the following section as bi-particles.

In terms of the Pauli spin operators a general form of Eq.(2.4) is

$$\rho_z(n_z n_x n_y) \equiv \frac{1}{2}(I + n_z \sigma_z + n_x \sigma_x + i n_y \sigma_y) \tag{2.6}$$

where $(n_z, n_x, n_y) = \{\pm 1, \pm 1, \pm 1\}$, which gives a more compact expression[5] for Eq.(2.5)

$$\rho_{\Psi_{12}^-} = \frac{1}{8} \begin{bmatrix} \left[\rho_z^1(+1,+1,+1) \otimes \rho_z^2(-1,-1,+1) + \rho_z^1(+1,+1,+1)^\dagger \otimes \rho_z^2(-1,-1,+1)^\dagger \right] + \\ \left[\rho_z^1(-1,-1,+1) \otimes \rho_z^2(+1,+1,+1) + \rho_z^1(-1,-1,+1)^\dagger \otimes \rho_z^2(+1,+1,+1)^\dagger \right] + \\ \left[\rho_z^1(+1,-1,+1) \otimes \rho_z^2(-1,+1,+1) + \rho_z^1(+1,-1,+1)^\dagger \otimes \rho_z^2(-1,+1,+1)^\dagger \right] + \\ \left[\rho_z^1(-1,+1,+1) \otimes \rho_z^2(+1,-1,+1) + \rho_z^1(-1,+1,+1)^\dagger \otimes \rho_z^2(+1,-1,+1)^\dagger \right] \end{bmatrix} \tag{2.7}$$

where $\dagger$ denotes the hermitian conjugate operation. In fact all four Bell states

$$|\psi_{12}^\pm\rangle = \frac{1}{\sqrt{2}}\left(|+1\rangle_z^1|-1\rangle_z^2 \pm |-1\rangle_z^1|+1\rangle_z^2\right) \quad |\phi_{12}^\pm\rangle = \frac{1}{\sqrt{2}}\left(|+1\rangle_z^1|+1\rangle_z^2 \pm |-1\rangle_z^1|-1\rangle_z^2\right) \tag{2.8}$$

can be decomposed into a sum of products states, and the other three are[5],

$$\rho_{\Psi_{12}^+} = \frac{1}{8} \begin{bmatrix} \left[\rho_z^1(+1,+1,+1) \otimes \rho_z^2(-1,+1,+1)^\dagger + \rho_z^1(+1,+1,+1) \otimes \rho_z^2(-1,+1,+1)^\dagger \right] + \\ \left[\rho_z^1(-1,-1,+1) \otimes \rho_z^2(+1,-1,+1)^\dagger + \rho_z^1(-1,-1,+1) \otimes \rho_z^2(+1,-1,+1)^\dagger \right] + \\ \left[\rho_z^1(+1,-1,+1) \otimes \rho_z^2(-1,-1,+1)^\dagger + \rho_z^1(+1,-1,+1) \otimes \rho_z^2(-1,-1,+1)^\dagger \right] + \\ \left[\rho_z^1(-1,+1,+1) \otimes \rho_z^2(+1,+1,+1)^\dagger + \rho_z^1(-1,+1,+1) \otimes \rho_z^2(+1,+1,+1)^\dagger \right] \end{bmatrix} \tag{2.9}$$

$$\rho_{\phi_{12}^-} = \frac{1}{8} \begin{bmatrix} \left[\rho_z^1(+1,+1,+1) \otimes \rho_z^2(+1,-1,+1)^\dagger + \rho_z^1(+1,+1,+1)^\dagger \otimes \rho_z^2(+1,-1,+1) \right] + \\ \left[\rho_z^1(-1,-1,+1) \otimes \rho_z^2(-1,+1,+1)^\dagger + \rho_z^1(-1,-1,+1)^\dagger \otimes \rho_z^2(-1,+1,+1) \right] + \\ \left[\rho_z^1(+1,-1,+1) \otimes \rho_z^2(+1,+1,+1)^\dagger + \rho_z^1(+1,-1,+1)^\dagger \otimes \rho_z^2(+1,+1,+1) \right] + \\ \left[\rho_z^1(-1,+1,+1) \otimes \rho_z^2(-1,-1,+1)^\dagger + \rho_z^1(-1,+1,+1)^\dagger \otimes \rho_z^2(-1,-1,+1) \right] \end{bmatrix} \tag{2.10}$$

And

$$\rho_{\phi_{12}^+} = \frac{1}{8} \begin{bmatrix} \left[\rho_z^1(+1,+1,+1) \otimes \rho_z^2(+1,+1,+1) + \rho_z^1(+1,+1,+1)^\dagger \otimes \rho_z^2(+1,+1,+1)^\dagger \right] + \\ \left[\rho_z^1(-1,-1,+1) \otimes \rho_z^2(-1,-1,+1) + \rho_z^1(-1,-1,+1)^\dagger \otimes \rho_z^2(-1,-1,+1)^\dagger \right] + \\ \left[\rho_z^1(+1,-1,+1) \otimes \rho_z^2(+1,-1,+1) + \rho_z^1(+1,-1,+1)^\dagger \otimes \rho_z^2(+1,-1,+1)^\dagger \right] + \\ \left[\rho_z^1(-1,+1,+1) \otimes \rho_z^2(-1,+1,+1) + \rho_z^1(-1,+1,+1)^\dagger \otimes \rho_z^2(-1,+1,+1)^\dagger \right] \end{bmatrix} \tag{2.11}$$

For completeness the statistical operators for the four Bell states are expressed in terms of the Pauli spin operators[5],

$$\rho_{\psi_{12}^-} = |\psi_{12}^-\rangle\langle\psi_{12}^-| = \frac{1}{4}(I^1 \otimes I^2 - \boldsymbol{\sigma}^1 \cdot \boldsymbol{\sigma}^2) \tag{2.12}$$

$$\rho_{\psi_{12}^+} = |\psi_{12}^+\rangle\langle\psi_{12}^+| = \frac{1}{4}(I^1 \otimes I^2 + \boldsymbol{\sigma}^1 \cdot \boldsymbol{\sigma}^2 - 2\sigma_z^1 \otimes \sigma_z^2) \tag{2.13}$$

$$\rho_{\Phi_{12}^-} = |\phi_{12}^-\rangle\langle\phi_{12}^-| = \frac{1}{4}(I^1 \otimes I^2 + \boldsymbol{\sigma}^1 \cdot \boldsymbol{\sigma}^2 - 2\sigma_x^1 \otimes \sigma_x^2) \tag{2.14}$$

$$\rho_{\Phi_{12}^+} = |\phi_{12}^+\rangle\langle\phi_{12}^+| = \frac{1}{4}\left(I^1 \otimes I^2 + \boldsymbol{\sigma}^1 \cdot \boldsymbol{\sigma}^2 - 2\sigma_y^1 \otimes \sigma_y^2\right) \qquad (2.15)$$

For example the equivalence between, Eq.(2.12) and Eq.(2.2) follows by substituting the Pauli spin matrices,

$$\rho_{\psi_{12}^-} = \frac{1}{4}\left(I^1 \otimes I^2 - \boldsymbol{\sigma}^1 \cdot \boldsymbol{\sigma}^2\right) = \frac{1}{4}\left(I^1 \otimes I^2 - \sigma_x^1 \otimes \sigma_x^2 - \sigma_y^1 \otimes \sigma_y^2 - \sigma_z^1 \otimes \sigma_z^2\right) \qquad (2.16)$$

and likewise for the others.

### 3. SOME PROPERTIES OF THE NON-HERMITIAN STATES

The non-hermitian states, Eq.(2.6) have the same real eigenvalues as the statistical state operators Eq.(2.3). Moreover Eqs.(2.3) and (2.6) are idempotent so they both satisfy the condition for pure states. A difference arises between the eigenstates of each. As prepared for experiment, and after many particles have passed the Stern-Gerlach filter, the eigenvalues of $\sigma_Z$ are +1 and -1 with orthogonal quantum states in the laboratory frame of

$$|+\rangle_Z = \begin{pmatrix} 1 \\ 0 \end{pmatrix}_Z \text{ for } +1 \text{ and } |-\rangle_Z = \begin{pmatrix} 0 \\ 1 \end{pmatrix}_Z \text{ for } -1 \qquad (3.1)$$

In contrast the states of the single events before reaching the Stern-Gerlach filters are proposed to be fundamentally non-hermitian, Eq.(2.6). There are eight different choices of integers in Eq.(2.6) each giving eigenvalues of ±1 and with states that differ from the others only in sign: one such pair is obtained from the first octant of the body frame $(n_z, n_x, n_y) = \{1,1,1\}$

$$\sigma_z + \sigma_x + i\sigma_y = \begin{pmatrix} 1 & 2 \\ 0 & -1 \end{pmatrix}_z \qquad (3.2)$$

leading to non-orthogonal eigenstates[5],

$$|+\rangle_z = \begin{pmatrix} 1 \\ 0 \end{pmatrix}_z \text{ for } +1 \text{ and } |-\rangle_x = \frac{1}{\sqrt{2}}\begin{pmatrix} 1 \\ -1 \end{pmatrix}_x \text{ for } -1 \qquad (3.3)$$

One eigenvalue is associated with the $\hat{\mathbf{z}}$ axis and the other with the $\hat{\mathbf{x}}$ axis of the body frame. For all choices of the integers in Eq.(2.6), the eigenvalues are always +1 and -1, and the associated eigenstates pairs for each octant are the various non-orthogonal combinations of[5]

$$\begin{pmatrix} 1 \\ 0 \end{pmatrix}_z ; \begin{pmatrix} 0 \\ 1 \end{pmatrix}_z ; \frac{1}{\sqrt{2}}\begin{pmatrix} 1 \\ -1 \end{pmatrix}_z ; \frac{1}{\sqrt{2}}\begin{pmatrix} 1 \\ 1 \end{pmatrix}_z . \qquad (3.4)$$

That is the eigenvalues of ±1 are eight-fold degenerate. There are no eigenstates associated with $y$ component.

The laboratory states, $|\pm\rangle_Z^i$, cannot be defined by the single event of one spin passing a filter. Rather the states are defined, along with their probabilities, after a statistically large number have been filtered[6]. In contrast the body frame states $|\pm\rangle_z^i$ and $|\pm\rangle_x^i$ do not represent ensembles, as the states $|\pm\rangle_Z^i$ do, but those of single particles. Further discussion of the statistical relationship between the states in the laboratory and body frames is deferred to next section.

In an isotropic environment there is no difference between the $\hat{\mathbf{z}}$ and $\hat{\mathbf{x}}$ quantization axes in Eq.(2.6), making these two components indistinguishable. Therefore the coherence in Eq.(2.6) leads to a hermitian resonance state defined by[5],

$$\rho_{z,\sqrt{2}}\left(\hat{\mathbf{n}}_{n_z n_x}\right) = \frac{1}{2}\left(\rho_z\left(n_z n_x n_y\right) + \rho_z\left(n_z n_x n_y\right)^\dagger\right) = \frac{1}{2}\left(I + \sqrt{2}\boldsymbol{\sigma} \cdot \hat{\mathbf{n}}_{n_z n_x}\right) \qquad (3.5)$$

The unit vectors in Eq.(3.5) bisects the four indistinguishable axes of the body frame $(n_z, n_x) = \{\pm 1, \pm 1\}$,

$$\hat{\mathbf{n}}_{n_z n_x} = \frac{1}{\sqrt{2}}\left(n_z \hat{\mathbf{z}} + n_x \hat{\mathbf{x}}\right) \qquad (3.6)$$

while the anti-hermitian difference is given by a rotation operator[5],

$$\frac{1}{2}\left(\rho_z\left(n_z n_x n_y\right) - \rho_z\left(n_z n_x n_y\right)^\dagger\right) = \frac{1}{2}in_y\sigma_y = \pm\frac{i\sigma_y}{2} \tag{3.7}$$

The hermitian resonance state, Eq.(3.5) defines a new spin with eigenvalues of $\pm\sqrt{2}$ ($\pm 1/\sqrt{2}$) rather that $\pm 1$ ($\pm\frac{1}{2}$). From Eq.(3.5) two sets of orthogonal states for $\sigma_z + \sigma_x$ are obtained[5],

$$\begin{pmatrix} 1 & 1 \\ 1 & -1 \end{pmatrix}_z \text{ gives eigenvalues of } \pm\sqrt{2} \quad \text{and eigenstates of} \quad |+\rangle^{\sqrt{2}}_{\pm 1,\pm 1,z} = \frac{1}{\sqrt{2(2+\sqrt{2})}}\begin{pmatrix} 1+\sqrt{2} \\ 1 \end{pmatrix}_z \tag{3.8}$$

$$|-\rangle^{\sqrt{2}}_{\pm 1,\pm 1,z} = \frac{1}{\sqrt{2(2+\sqrt{2})}}\begin{pmatrix} 1 \\ -1-\sqrt{2} \end{pmatrix}_z$$

and two sets of orthogonal states for $\sigma_z - \sigma_x$

$$\begin{pmatrix} 1 & -1 \\ -1 & -1 \end{pmatrix}_z \text{ gives eigenvalues of } \pm\sqrt{2} \quad \text{and eigenstates of} \quad |+\rangle^{\sqrt{2}}_{\pm,1\mp,1,z} = \frac{1}{\sqrt{2(2+\sqrt{2})}}\begin{pmatrix} -1-\sqrt{2} \\ 1 \end{pmatrix}_z \tag{3.9}$$

$$|-\rangle^{\sqrt{2}}_{\pm,1\mp,1,z} = \frac{1}{\sqrt{2(2+\sqrt{2})}}\begin{pmatrix} 1 \\ 1+\sqrt{2} \end{pmatrix}_z$$

Geometrically these eigenstates lie at angles of 45° between the four sets of the two axes defined by $n_z\hat{\mathbf{z}}$ and $n_x\hat{\mathbf{x}}$ and are a superposition of the non-orthogonal states in Eq.(3.3). The resulting resonance spin of magnitude $\sqrt{2}$ is purely quantum in origin. In terms of the usual spin magnitude of $\hbar/2$, the operator for the above spin states for a single particle is defined by,

$$\mathbf{S}_{\sqrt{2}} \equiv \frac{\hbar}{\sqrt{2}}\boldsymbol{\sigma} \tag{3.10}$$

If the environment is not isotropic, such as when the spin interacts with local or external magnetic fields, the two axes are distinguishable and the $\sqrt{2}$ resonance state cannot be formed. In that case the spin is described by the non-hermitian state, Eq.(2.6). The spin, Eq.(3.10) is predicted to exist in an isotropic environment only.

## 4. INTERPRETATION

Whereas entanglement, observed in the Bell states, evaporates when more degrees of freedom are introduced as seen from Eq.(2.6), the question arises as to whether these non-hermitian operators represent the states of actual particles, or whether the separation of the Bell states is a mathematical property of the Pauli spin operators.

Consider first an isolated spin ½ as described from Eqs.(3.5) to (3.10). In its body frame, the $\sqrt{2}$ spin can be oriented in any of four ways depending on $\hat{\mathbf{n}}_{n_z n_x}$, Eq.(3.6), so the eigenvalues of each, Eqs.(3.8) and (3.9), are four-fold degenerate. These forms a resonance state of overall zero angular momentum consistent with a point particle. In the presence of an external magnetic field the two axes are no longer indistinguishable and the resonance spin is destroyed. In this case one or the other body axes aligns with the external field and the other component is randomized away. This restores the usual spin ½ as defined by a Stern-Gerlach filter, Eq.(2.3). In this section these conclusions are discussed.

The singlet is also found to be a resonance state between four indistinguishable orientations of the two spins in the body frame. As expected, each is composed of two spins oppositely aligned along a common axis. There is no net angular momentum in any orientation, so each is equally probable. These resonance states represent bi-particles and their sum gives the singlet state. The two spins must be close enough to oppositely align, thereby justifying the distinction that this is a bound or local singlet. It is composed of two spins only and they share the same body frame.

A singlet can also be expressed by ensemble averaging over bi-particles formed from the decomposition of many bound singlet states. At the instant a bound singlet decomposes, it does so from one of its resonance states to produces one bi-particle from the four possible. Each particle then moves in opposite directions: one to Alice and the other to Bob and can no longer influence each other. However their two body frames are identical.

Bi-particles share a second rank tensor correlation[3] and this is responsible for the apparent entanglement at space-like separations. The second rank tensor contains a rotationally invariant part which is identical for all bi-particles even though the body frames differ from one singlet to another. The ensemble average over all these bi-particles leaves only this invariant part which is mathematically identical to the bound singlet. Over a statistical measurement, the invariant correlation builds up and leads to the violation of Bell's Inequalities[7]. Whereas the bound singlet is an exact representation for two spins, in contrast the ensemble average over many particles removes anisotropic terms and is therefore not exact. To distinguish this from the bound singlet, it is termed the unbound or non-local singlet. This has nothing to do with non-local connectivity[8]. In this treatment, Einstein locality[9] is assumed. These cases are discussed below.

## *Spin ½ structure and resonance*.

From Eq.(3.3) the eigenstates are associated with the angular momentum in both the body fixed $\hat{\mathbf{z}}$ and $\hat{\mathbf{x}}$ axes. This endows a two dimensional structure on a single spin. The non-hermiticity is a result of the term, $i\sigma_y = \sigma_z \sigma_x$ which orients the two quantization axes in 3D space. An orientation cannot be directly measured and a rotation operator is not an observable: recall there are no states associated with the *y* axis. Opposite phases (180° apart) of the 2D spin, Eq.(3.7), differentiate the two indistinguishable orientations.

In spite of this structure, in the absence of external interactions, the states, Eqs.(3.8) and (3.9), are eight-fold degenerate, leading to four resonances such that the angular momentum in the $\hat{\mathbf{z}}$, $\hat{\mathbf{x}}$ plane cancels,

$$\rho_{z,\sqrt{2}} = \frac{1}{4}\left(\rho_{z,\sqrt{2}}\left(\hat{\mathbf{n}}_{11}\right) + \rho_{z,\sqrt{2}}\left(\hat{\mathbf{n}}_{-1-1}\right) + \rho_{z,\sqrt{2}}\left(\hat{\mathbf{n}}_{1-1}\right) + \rho_{z,\sqrt{2}}\left(\hat{\mathbf{n}}_{-11}\right)\right) = \frac{I}{2} \qquad (4.1)$$

This is consistent with the conclusion that an electron is a point particle but one with zero net angular momentum. As shown below, in the presence of external interactions, only one of the resonances survives and the usual spin with magnitude ½ emerges.

A wave vector is insufficient to describe the coherences that lead to these resonances. Rather the coherence must be described by an operator, see *e.g.* Eq.(2.4) which permits off-diagonal elements. Even so these have the properties of pure states, being idempotent with unit trace. The time dependence is given by the quantum Liouville equation rather than the Schrödinger equation.

The above interpretation and the intuitively reasonable idea of separability of particles requires that quantum mechanics be extended to allow for non-hermitian states, and identifies these as the pure states of single particles that are normally identified as spin ½. In spite of this, in the cases treated, the non-hermiticity is concealed when the spins are either isolated or interacting. Isolated non-hermitian states form √2 hermitian spins; or they combine with other spins when in close proximity to form, for example, the hermitian Bell states.

## *Bi-particles and the bound resonance singlet.*

The singlet state in Eq.(2.7) has eight terms and these are grouped together as four, each of which is hermitian and anisotropic with respect to the four vectors[5], $\hat{\mathbf{n}}_{n_z,n_x}$, Eq.(3.6),

$$\rho_{\psi_{\bar{1}2}}\left(\hat{\mathbf{n}}_{n_z,n_x}\right) \equiv \frac{1}{2}\left[\rho_z^1\left(n_z,n_x,1\right) \otimes \rho_z^2\left(-n_z,-n_x,1\right) + \rho_z^1\left(n_z,n_x,1\right)^\dagger \otimes \rho_z^2\left(-n_z,-n_x,1\right)^\dagger\right] \qquad (4.2)$$

The vectors are opposite for each spin and are therefore 'paired'. The hermitian conjugate is due to the two axes being indistinguishable in the body frame. Equation (4.2) is a product state and therefore not entangled. The four anisotropic terms in Eq.(2.7) sum to give the isotropic, resonant and bound singlet,

$$\rho_{\Psi_{\bar{1}2}} = \frac{1}{4}\left[\rho_{\psi_{\bar{1}2}}\left(\hat{\mathbf{n}}_{11}\right) + \rho_{\psi_{\bar{1}2}}\left(\hat{\mathbf{n}}_{-1-1}\right) + \rho_{\psi_{\bar{1}2}}\left(\hat{\mathbf{n}}_{1-1}\right) + \rho_{\psi_{\bar{1}2}}\left(\hat{\mathbf{n}}_{-11}\right)\right] = \frac{1}{4}\left(I^1 \otimes I^2 - \boldsymbol{\sigma}^1 \cdot \boldsymbol{\sigma}^2\right) \qquad (4.3)$$

Since the eigenvalues are the same for each, and the eigenstates differ in sign only, the singlet as expressed in Eq.(4.3) is a resonance state between four degenerate orientations (*i.e.* different quadrants of the body frame). Each term is called a bi-particle partly because it is not a particle in the sense that it is not pure, and a bi-particle is one of the four degenerate contributions to the singlet state of two spins.

By definition resonance means that at any instant the pair is in one of its four degenerate orientations. When two spins are close enough to interact, the four terms exist with equal probability and the anisotropy of each bi-particle cancels leaving the bound singlet state (substitute Eq.(2.6) into Eq.(4.2), and then into Eq.(4.3).)

Each term in Eq.(4.2) can be expressed as

$$\rho_z(n_z n_x n_y) = \frac{1}{2}\left(I + \sqrt{3}\boldsymbol{\sigma} \cdot \hat{\mathbf{n}}_{n_z n_x n_y}\right) \quad (4.4)$$

with a complex unit vector,

$$\hat{\mathbf{n}}_{n_z n_x n_y} = \frac{1}{\sqrt{3}}\left(n_z \hat{\mathbf{z}} + n_x \hat{\mathbf{x}} + i n_y \hat{\mathbf{y}}\right) \quad (4.5)$$

The norm is defined by the scalar product[10], $\hat{\mathbf{n}}^* \cdot \hat{\mathbf{n}} = |\hat{\mathbf{n}}|^2 = 1$, so that the hermitian conjugate Eq.(4.4) is

$$\rho_z^\dagger(n_z n_x n_y) \equiv \frac{1}{2}\left(I + \sqrt{3}\boldsymbol{\sigma} \cdot \hat{\mathbf{n}}^*_{n_z n_x n_y}\right) = \rho_z(n_z n_x -n_y) \quad (4.6)$$

Substitution of Eq.(4.4) into Eq.(4.2) leads to the state for each bi-particle which carries a √2 spin aligned oppositely from each other[5],

$$\rho_{\psi_{\bar{12}}}(\hat{\mathbf{n}}_{n_z,n_x}) = \frac{1}{4}\left[I^1 \otimes I^2 + \sqrt{2}\boldsymbol{\sigma}^1 \cdot \hat{\mathbf{n}}_{n_z n_x} \otimes I^2 - I^1 \otimes \sqrt{2}\boldsymbol{\sigma}^2 \cdot \hat{\mathbf{n}}_{n_z n_x} - \frac{3}{2}\boldsymbol{\sigma}^1 \cdot \left(\hat{\mathbf{n}}_{n_z n_x 1}\hat{\mathbf{n}}^*_{n_z n_x 1} + \hat{\mathbf{n}}^*_{n_z n_x 1}\hat{\mathbf{n}}_{n_z n_x 1}\right) \cdot \boldsymbol{\sigma}^2\right] \quad (4.7)$$

Summing the four terms in Eq.(4.7) and noting that[5],

$$\sum_{n_z n_x =-1}^{1} \hat{\mathbf{n}}_{n_z n_x} = 0$$

$$\sum_{n_z n_x =-1}^{1} \left(\hat{\mathbf{n}}_{n_z n_x 1}\hat{\mathbf{n}}^*_{n_z n_x 1} + \hat{\mathbf{n}}^*_{n_z n_x 1}\hat{\mathbf{n}}_{n_z n_x 1}\right) = \frac{2}{3}\underline{\underline{\mathbf{U}}} \quad (4.8)$$

shows that only the rotationally invariant parts remain giving the bound singlet as,

$$\rho_{\psi_{\bar{12}}} = \frac{1}{4}\left[I^1 \otimes I^2 + \sqrt{2}\boldsymbol{\sigma}^1 \cdot \sum_{n_z n_x =-1}^{1} \hat{\mathbf{n}}_{n_z n_x} \otimes I - I^1 \otimes \sqrt{2}\boldsymbol{\sigma}^2 \cdot \sum_{n_z n_x =-1}^{1} \hat{\mathbf{n}}_{n_z n_x} - \frac{3}{2}\boldsymbol{\sigma}^1 \cdot \sum_{n_z n_x =-1}^{1} \left(\hat{\mathbf{n}}_{n_z n_x 1}\hat{\mathbf{n}}^*_{n_z n_x 1} + \hat{\mathbf{n}}^*_{n_z n_x 1}\hat{\mathbf{n}}_{n_z n_x 1}\right) \cdot \boldsymbol{\sigma}^2\right]$$
$$= \frac{1}{4}\left[I^1 \otimes I^2 - \boldsymbol{\sigma}^1 \cdot \underline{\underline{\mathbf{U}}} \cdot \boldsymbol{\sigma}^2\right] = \frac{1}{4}\left[I^1 \otimes I^2 - \boldsymbol{\sigma}^1 \cdot \boldsymbol{\sigma}^2\right] \quad (4.9)$$

Here the completely symmetric second rank tensor is $\underline{\underline{\mathbf{U}}} = \hat{\mathbf{z}}^2 + \hat{\mathbf{x}}^2 + \hat{\mathbf{y}}^2$.

## *Bi-particles and the unbound correlated singlet.*

In its bound resonance state, all four bi-particles contribute equally and Eq.(4.3) is an exact expression for a bound singlet state, Eq.(2.12). At the instant the two spins separate, only one of the four bi-particles forms. In this section, each hermitian bi-particle is shown to correspond to two separated and non-interacting particles that are correlated due to their common body axis. Although the vector spins in Eq.(4.7) cancel, within the bound singlet they are viable after separation. The single particle states are obtained by tracing over one spin[5],

$$\rho^1_{z,\sqrt{2}}(\mathbf{n}_{n_z n_x}) = \text{Tr}_2\left[\rho_{\psi_{\bar{12}}}(\hat{\mathbf{n}}_{n_z,n_x})\right] = \frac{1}{2}\left[I^1 + \sqrt{2}\boldsymbol{\sigma}^1 \cdot \hat{\mathbf{n}}_{n_z n_x}\right] \quad (4.10)$$

which is valid only in an isotropic environment after separation: spin 1 moves towards Alice and an identical spin with $\hat{\mathbf{n}}_{n_z n_x} \to -\hat{\mathbf{n}}_{n_z n_x}$, moves towards Bob,

$$\rho^2_{z,\sqrt{2}}(\mathbf{n}_{n_z n_x}) = \text{Tr}_1\left[\rho_{\psi_{\bar{12}}}(\hat{\mathbf{n}}_{n_z,n_x})\right] = \frac{1}{2}\left[I^2 - \sqrt{2}\boldsymbol{\sigma}^2 \cdot \hat{\mathbf{n}}_{n_z n_x}\right] \quad (4.11)$$

Therefore, over space-like separations the two spins remain antiparallel and correlate. On the other hand, if the source produces randomly oriented spins, ensemble averaging renders the vector spins to zero. For this reason, they cannot account for the correlation that leads to violation of Bell's Inequalities[7]. However each bi-particle is also correlated to its partner by virtue of the persistence of the rotationally invariant contribution, $\boldsymbol{\sigma}^1 \cdot \boldsymbol{\sigma}^2$. It is stressed that each bi-particle, also identified as EPR[9] pairs, carries the same invariant term that survives ensemble averaging. It is a fundamentally dilemma of quantum theory as to how the quantum correlation is maintained over space-like separations. In terms of this treatment, with the assumption of Einstein locality[9], the rotationally invariant term, $\boldsymbol{\sigma}^1 \cdot \boldsymbol{\sigma}^2$ is the same for all body frame axes,

$$\boldsymbol{\sigma}^1 \cdot \boldsymbol{\sigma}^2 = 3\boldsymbol{\sigma}^1 \cdot \hat{\mathbf{n}}\hat{\mathbf{n}}^* \cdot \boldsymbol{\sigma}^2 \quad (4.12)$$

where the unit vector is defined by Eq.(4.5). Every bi-particle has a similar form but with differing $\hat{\mathbf{n}}$. As long as they originated from the same bound singlet, and no interactions have intervened, then Eq.(4.12) is maintained. Hence if one spin, say Eq.(4.10), is filtered, then in coincidence experiments on photons[11,12] its distant partner will show strong correlation if the filter are set to correspond to angular momentum along the same body axes as its partner. Suppose the filters in these experiments are set at $\hat{\mathbf{a}}$ and $\hat{\mathbf{b}}$, then using Eq.(4.7) the expectation value for one bi-particle is[5],

$$\left\langle \hat{\mathbf{a}} \cdot \boldsymbol{\sigma}^1 \boldsymbol{\sigma}^2 \cdot \hat{\mathbf{b}} \right\rangle = \hat{\mathbf{a}} \cdot \operatorname*{Tr}_{1,2}\left[ \rho_{\psi_{12}^-}\left( \hat{\mathbf{n}}_{n_z, \pm n_x} \right) \boldsymbol{\sigma}^1 \boldsymbol{\sigma}^2 \right] \cdot \hat{\mathbf{b}}$$
$$= -\cos\theta_{ab} \pm n_z n_x \left( \delta(\hat{\mathbf{a}}-\hat{\mathbf{z}})\delta(\hat{\mathbf{b}}-\hat{\mathbf{x}}) \pm \delta(\hat{\mathbf{a}}-\hat{\mathbf{x}})\delta(\hat{\mathbf{b}}-\hat{\mathbf{z}}) \right) \cong -\cos\theta_{ab} \quad (4.13)$$

It is stressed that this expression describes the coincidence correlation between only two particles $3\hat{\mathbf{a}}\cdot\hat{\mathbf{n}}\hat{\mathbf{n}}^*\cdot\hat{\mathbf{b}} = \hat{\mathbf{a}}\cdot\hat{\mathbf{b}} = \cos\theta_{ab}$ and is responsible for the violation of Bell's inequalities[8]. The anisotropic terms in Eq.(4.13) are unlikely to survive statistical measurements unless the ensemble lacks isotropy[13]. Note again the bi-particle state, Eq.(4.7), is separable as seen from the product form, Eq.(4.2).

Similar expressions are obtained for the other three Bell states which are given in terms of sums of four similar bi-particles which differ only in the way they combine, Eqs.(4.31) to (4.33). In all cases, the bi-particles display a scalar part and two vectors and a second rank tensor[3]. These have antisymmetric parts and a rotationally invariant trace and differ for the four Bell states, (cf. Eq.(2.12) to (2.15)[5],

$$\begin{aligned}
\text{For } \psi_{12}^- \quad & \left( \hat{\mathbf{n}}_{n_z n_x 1}\hat{\mathbf{n}}^*_{n_z n_x 1} + \hat{\mathbf{n}}^*_{n_z n_x 1}\hat{\mathbf{n}}_{n_z n_x 1} \right) = \frac{2}{3}\left[ \underline{\underline{\mathbf{U}}} + n_z n_x \left( \hat{\mathbf{z}}\hat{\mathbf{x}} + \hat{\mathbf{x}}\hat{\mathbf{z}} \right) \right] \\
\text{For } \psi_{12}^+ \quad & \left( \hat{\mathbf{n}}_{n_z n_x 1}\hat{\mathbf{n}}^*_{-n_z n_x 1} + \hat{\mathbf{n}}^*_{n_z n_x 1}\hat{\mathbf{n}}_{-n_z n_x 1} \right) = \frac{2}{3}\left[ \underline{\underline{\mathbf{U}}} - 2\hat{\mathbf{z}}^2 + n_z n_x \left( \hat{\mathbf{z}}\hat{\mathbf{x}} - \hat{\mathbf{x}}\hat{\mathbf{z}} \right) \right] \\
\text{For } \phi_{12}^- \quad & \left( \hat{\mathbf{n}}_{n_z n_x 1}\hat{\mathbf{n}}^*_{n_z -n_x 1} + \hat{\mathbf{n}}^*_{n_z n_x 1}\hat{\mathbf{n}}_{n_z -n_x 1} \right) = \frac{2}{3}\left[ \underline{\underline{\mathbf{U}}} - 2\hat{\mathbf{x}}^2 - n_z n_x \left( \hat{\mathbf{z}}\hat{\mathbf{x}} - \hat{\mathbf{x}}\hat{\mathbf{z}} \right) \right] \\
\text{For } \phi_{12}^+ \quad & \left( \hat{\mathbf{n}}_{n_z n_x 1}\hat{\mathbf{n}}_{n_z n_x 1} + \hat{\mathbf{n}}^*_{n_z n_x 1}\hat{\mathbf{n}}^*_{n_z n_x 1} \right) = \frac{2}{3}\left[ \underline{\underline{\mathbf{U}}} - 2\hat{\mathbf{y}}^2 + n_z n_x \left( \hat{\mathbf{z}}\hat{\mathbf{x}} + \hat{\mathbf{x}}\hat{\mathbf{z}} \right) \right]
\end{aligned} \quad (4.14)$$

Bi-particles emitted from the same bound Bell state contain invariant contributions. The antisymmetric terms $n_z n_x \left( \hat{\mathbf{z}}\hat{\mathbf{x}} \pm \hat{\mathbf{x}}\hat{\mathbf{z}} \right)$ vary in sign over the range of $n_z n_x n_y$ and all exactly cancel when summed, see Eq.(4.8). For an ensemble measurement over a random distribution of $\hat{\mathbf{n}}$ only the rotationally invariant part, $\underline{\underline{\mathbf{U}}}$ survives which leads to the unbound singlet[5],

$$\rho_{\psi_{12}^-} = \overline{\rho_{\psi_{12}^-}\left( \mathbf{n}_{n_z n_x} \right)}^{EA} = \frac{1}{4}\left( I^1 \otimes I^2 - \boldsymbol{\sigma}^1 \cdot \boldsymbol{\sigma}^2 \right) \quad (4.15)$$

rendering an expression which is identical to the bound singlet state, Eq.(4.9). Although the bound and unbound singlets are mathematically equivalent, Eq.(2.12), their origin is different. The bound singlet is an exact resonance between two particles, Eq.(4.3) while the unbound singlet is an approximation due to the ensemble average. Similar results are found for the other Bell states which are given at the end of this section.

In summary, a singlet state takes the same mathematical form in both bound and unbound cases but describe different situations:

The bound case treats two spins that are close enough so their respective fields interact. Although randomly oriented in 3D space, the two spins share the same body frame and form four degenerate or resonance states as shown in Eq.(4.3). Each resonance is a product state called a bi-particle. Summing over the four vectors, Eq.(4.5) using Eq.(4.8), leaves only the rotationally invariant parts and gives the bound singlet state.

The unbound case corresponds to two separated spins that obey Einstein locality[9] and retain correlation between them as long their body frames coincide. As one singlet separates, it can do so only into one of its four bi-particle states and carries both anisotropic and isotropic terms that are the same as in their original bound singlet, Eq.(4.7). Ensemble averaging over a random distribution of bi-particles leaves only the rotationally invariant part which is identical for every bi-particle and is mathematically indistinguishable from the bound singlet.

## *Non-hermitian spins in the laboratory frame and dechoherence*.

The non-hermitian states, *e.g.* Eq.(2.4), form √2 spins and these lead to resonant point particles in the absence of external interactions. In the presence of a second spin, the two particle (Bell) states form. In the transition between the two, or in the presence of any interaction strong or close enough to render the two body axes distinguishable, the non-hermitian state exists. In this section the spins oriented in their body frame are transformed to the laboratory frame where it is shown that decoherence renders the macroscopic states hermitian, and leads to consistency with the usual results from quantum theory.

The laboratory frame is chosen randomly in any convenient direction in 3D space. Although an interaction can render space anisotropic, for the purposes of this section assume only local interactions act within the proximity of the spin. This can be viewed as a test particle so the overall space remains isotropic. The body and the laboratory frames differ by a rotation by[5] $\theta, \phi$

$$\hat{\mathbf{z}} = \cos\theta \hat{\mathbf{Z}} + \sin\theta \cos\varphi \hat{\mathbf{X}} + \sin\theta \sin\varphi \hat{\mathbf{Y}}$$
$$\hat{\mathbf{x}} = -\sin\theta \hat{\mathbf{Z}} + \cos\theta \cos\varphi \hat{\mathbf{X}} + \cos\theta \sin\varphi \hat{\mathbf{Y}} \quad (4.16)$$
$$\hat{\mathbf{y}} = -\sin\varphi \hat{\mathbf{X}} + \cos\varphi \hat{\mathbf{Y}}$$

The states in the body and laboratory frames are related by[5]

$$|+\rangle_z = \begin{pmatrix} \cos\theta/2 \, e^{-i\varphi/2} \\ \sin\theta/2 \, e^{i\varphi/2} \end{pmatrix}_Z \quad |-\rangle_x = \frac{1}{\sqrt{2}} \begin{pmatrix} -(\cos(\theta/2) + \sin(\theta/2)) e^{-i\varphi/2} \\ (\cos(\theta/2) - \sin(\theta/2)) e^{+i\varphi/2} \end{pmatrix}_Z$$
$$|-\rangle_z = \begin{pmatrix} -\sin(\theta/2) e^{-i\varphi/2} \\ \cos(\theta/2) e^{+i\varphi/2} \end{pmatrix}_Z \quad |+\rangle_x = \frac{1}{\sqrt{2}} \begin{pmatrix} (\cos(\theta/2) - \sin(\theta/2)) e^{-i\varphi/2} \\ (\cos(\theta/2) + \sin(\theta/2)) e^{i\varphi/2} \end{pmatrix}_Z \quad (4.17)$$

and the Pauli spin components are expressed in terms of the raising and lowering operators $\sigma_\pm = \sigma_X \pm i\sigma_Y$ by[5]

$$\sigma_z = \cos\theta \sigma_Z + \frac{1}{2}\sin\theta \left( e^{-i\phi}\sigma_+ + e^{+i\phi}\sigma_- \right)$$
$$\sigma_x = -\sin\theta \sigma_Z + \frac{1}{2}\cos\theta \left( e^{-i\phi}\sigma_+ + e^{+i\phi}\sigma_- \right) \quad (4.18)$$
$$i\sigma_y = +\frac{1}{2}\left( e^{-i\phi}\sigma_+ - e^{+i\phi}\sigma_- \right)$$

Since transformation from any body frame octant to the laboratory frame differs only by signs, the treatment is restricted to the first octant: the others follow similarly. Substitution of Eqs.(4.18) into Eq.(2.4) gives[5]

$$\rho_z(111) \equiv \frac{1}{2}\left( I + \cos\theta \sigma_Z - \sin\theta \sigma_Z + \frac{1}{2}\left( e^{-i\phi}\sigma_+ (1+\sin\theta+\cos\theta) - e^{+i\phi}\sigma_- (1-\sin\theta-\cos\theta) \right) \right) \quad (4.19)$$

This represents the pure state of a single randomly oriented non-hermitian spin in the laboratory frame where the phase coherence, via $\phi$, is displayed. Random fluctuations cause phases to undergo decoherence[14], $\overline{e^{\pm i\phi}}^D = 0$ indicated by the overbar. This or integration over a distribution of many spins with random phase angles, leads directly to the hermitian statistical operator,

$$\rho_Z^D \equiv \overline{\rho_z(111)}^D = \frac{1}{2}\left( I + (\cos\theta - \sin\theta)\sigma_Z \right) \quad (4.20)$$

which is generally a mixed state with the trace of the square given by

$$\mathrm{Tr}\left[ \rho_Z^D \right]^2 = 1 - \cos\theta \sin\theta \quad (4.21)$$

Pure state are found in cases: $\theta = 0, \pi/2$. The cases, $\theta = 0$, correspond to an ensemble of spins with random phases all with the same body frame $\hat{\mathbf{z}}$ axis parallel or antiparallel with the laboratory axis ($\hat{\mathbf{Z}} = \pm\hat{\mathbf{z}}$). The eigenstates are given by $|\pm\rangle_z$, see Eq.(2.1). The second case, $\theta = \pi/2$, has opposite eigenvalues to the other pure case and corresponds to $\hat{\mathbf{Z}} = \pm\hat{\mathbf{x}}$. Since it is impossible to distinguish one body axis from the other, the states are the same, but opposite, respectively $|\mp\rangle_z$. Note that in the above development, the laboratory frame can be oriented in any direction because the space is assumed to be isotropic. The presence of a magnetic field destroys the isotropy and is treated next.

Finally for mixed states, $\theta \neq 0, \pi/2$, averaging over a random distribution of $\theta$ leads to the statistical operator for an unpolarized mixture,

$$\rho \equiv \overline{\rho_{x,y,z}}^{EA} = \frac{I}{2} \tag{4.22}$$

## *Non-hermitian spins in external fields.*

An[15] external field with major axis in the $\hat{\mathbf{Z}}$ direction destroys the isotropy of real space and causes the two body axes of each 2D spin to precess differently. Over an ensemble, all states randomize except those where one of the two body axes, $\hat{\mathbf{x}}$ or $\hat{\mathbf{z}}$, aligns with $\hat{\mathbf{Z}}$. In this calculation electron spin is used that can interact with either local moments or external magnetic fields. If a Stern-Gerlach filter is used, the magnetic field must be inhomogeneous in order to provide a coupling between the linear momentum and the spin angular momentum[16]. The objective here is not to show that beams are deflected up and down, but rather that one of the components aligns with the field while the others are randomized by ensemble averaging due to their time dependence in the field. A homogeneous field therefore suffices.

Although a completely isolated spin forms the √2 spin, this is destroyed as it approaches a magnetic field. Therefore the non-hermitian state is used in this calculation. Assume the external field is in the laboratory $\hat{\mathbf{Z}}$ direction, given by $\mathbf{H}_o = H_o \hat{\mathbf{Z}}$. Therefore the hamiltonian for the Zeeman interaction is,

$$H = -\boldsymbol{\mu}_e \cdot \hat{\mathbf{Z}} H_o \tag{4.23}$$

in terms of the magnetic moment of a spin, $\boldsymbol{\mu}_e$. Due to a single spin having two quantization axes, it is assumed a magnetic moment is associate with each, given in terms of the Bohr magneton, $\mu_\beta$, and the g-factor along each body frame axis,

$$\boldsymbol{\mu}_{ez} = -g\mu_\beta \sigma_z \hat{\mathbf{z}} \quad \text{and} \quad \boldsymbol{\mu}_{ex} = -g\mu_\beta \sigma_x \hat{\mathbf{x}} \tag{4.24}$$

There is no dipole-dipole interaction within a single electron because of the orthogonality of $\hat{\mathbf{x}}$ and $\hat{\mathbf{z}}$, so the magnetic moment is given by

$$\boldsymbol{\mu}_e = \boldsymbol{\mu}_{ez} - \boldsymbol{\mu}_{ex} = -g\mu_\beta \left( \sigma_z \hat{\mathbf{z}} - \sigma_x \hat{\mathbf{x}} \right) \tag{4.25}$$

Using the quantum Liouville equation,

$$i\hbar \frac{d\rho}{dt} = [H, \rho] \tag{4.26}$$

with the body frame state operator, Eq.(2.4) and the hamiltonian, Eq.(4.23) leads to

$$\frac{d(\boldsymbol{\mu}_{ez} - \boldsymbol{\mu}_{ex})}{dt} = -i\omega_o \left\{ (\boldsymbol{\mu}_{ex} \cos\theta + \boldsymbol{\mu}_{ez} \sin\theta) \right\} \tag{4.27}$$

with $\omega_o \equiv g\mu_\beta H_o$ and is independent of the phase angle $\phi$. This separates into two equations with solutions,

$$\frac{d\boldsymbol{\mu}_{ex}}{dt} = +i\cos\theta \omega_o \boldsymbol{\mu}_{ex} \to \boldsymbol{\mu}_{ex}(t) = e^{i\cos\theta \omega_o t} \boldsymbol{\mu}_{ex}(0)$$

$$\frac{d\boldsymbol{\mu}_{ez}}{dt} = -i\sin\theta \omega_o \boldsymbol{\mu}_{ez} \to \boldsymbol{\mu}_{ez}(t) = e^{-i\sin\theta \omega_o t} \boldsymbol{\mu}_{ez}(0) \tag{4.28}$$

showing that each body axis of quantization accumulates phase in opposite directions and with a frequency that is modulated by the angle the magnetic moments make with the external magnetic field.

This calculation shows that all states are averaged to zero except the component aligned with the field, hence for the states, $\theta = 0, \pi/2$ one or the other body axis is aligned with the field axis. For $\theta = 0$

$$\overline{\boldsymbol{\mu}_{ez}(t)}^{EA} = \boldsymbol{\mu}_{eZ} \qquad \overline{\boldsymbol{\mu}_{ex}(t)}^{EA} = 0 \tag{4.29}$$

or for $\theta = \pi/2$

$$\overline{\boldsymbol{\mu}_{ez}(t)}^{EA} = 0 \qquad \overline{\boldsymbol{\mu}_{ex}(t)}^{EA} = \boldsymbol{\mu}_{eZ} \tag{4.30}$$

The magnetic moment is defined in the laboratory axis, $\boldsymbol{\mu}_{eZ} = -g\mu_\beta \sigma_z \hat{\mathbf{Z}}$ in accord with the usual results from quantum theory and the pure states are given by Eqs.(2.3) or (3.1), with a single laboratory axis of quantization.

## *Bi-particle Bell states*

The treatment can be extended to the other Bell states which are summarized here. The singlet bi-particle equation (4.7) is repeated along with the other three[5]:

$$\rho_{\psi_{12}^-}\left(\hat{\mathbf{n}}_{n_z,n_x}\right) = \frac{1}{4}\left[I^1 \otimes I^2 + \sqrt{2}\boldsymbol{\sigma}^1 \cdot \hat{\mathbf{n}}_{n_z n_x} \otimes I^2 - I^1 \otimes \sqrt{2}\boldsymbol{\sigma}^2 \cdot \hat{\mathbf{n}}_{n_z n_x} - \boldsymbol{\sigma}^1 \cdot \left(\underline{\underline{U}} + n_z n_x \left(\hat{\mathbf{z}}\hat{\mathbf{x}} + \hat{\mathbf{x}}\hat{\mathbf{z}}\right)\right)\cdot\boldsymbol{\sigma}^2\right]$$
(4.7)

$$\rho_{\psi_{12}^+}\left(\hat{\mathbf{n}}_{n_z,n_x}\right) = \frac{1}{4}\left[I^1 \otimes I^2 + \sqrt{2}\boldsymbol{\sigma}^1 \cdot \hat{\mathbf{n}}_{n_z n_x} \otimes I^2 - I^1 \otimes \sqrt{2}\boldsymbol{\sigma}^2 \cdot \hat{\mathbf{n}}_{n_z, -n_x} + \boldsymbol{\sigma}^1 \cdot \left(\underline{\underline{U}} - 2\hat{\mathbf{z}}^2 + n_z n_x \left(\hat{\mathbf{z}}\hat{\mathbf{x}} - \hat{\mathbf{x}}\hat{\mathbf{z}}\right)\right)\cdot\boldsymbol{\sigma}^2\right]$$
(4.31)

$$\rho_{\phi_{12}^-}\left(\hat{\mathbf{n}}_{n_z,n_x}\right) = \frac{1}{4}\left[I^1 \otimes I^2 + \sqrt{2}\boldsymbol{\sigma}^1 \cdot \hat{\mathbf{n}}_{n_z n_x} \otimes I^2 + I^1 \otimes \sqrt{2}\boldsymbol{\sigma}^2 \cdot \hat{\mathbf{n}}_{n_z, -n_x} + \boldsymbol{\sigma}^1 \cdot \left(\underline{\underline{U}} - 2\hat{\mathbf{x}}^2 - n_z n_x \left(\hat{\mathbf{z}}\hat{\mathbf{x}} - \hat{\mathbf{x}}\hat{\mathbf{z}}\right)\right)\cdot\boldsymbol{\sigma}^2\right]$$
(4.32)

$$\rho_{\phi_{12}^+}\left(\hat{\mathbf{n}}_{n_z,n_x}\right) = \frac{1}{4}\left[I^1 \otimes I^2 + \sqrt{2}\boldsymbol{\sigma}^2 \cdot \hat{\mathbf{n}}_{n_z n_x} \otimes I^2 + I^1 \otimes \sqrt{2}\boldsymbol{\sigma}^2 \cdot \hat{\mathbf{n}}_{n_z n_x} + \boldsymbol{\sigma}^1 \cdot \left(\underline{\underline{U}} - 2\hat{\mathbf{y}}^2 + n_z n_x \left(\hat{\mathbf{z}}\hat{\mathbf{x}} + \hat{\mathbf{x}}\hat{\mathbf{z}}\right)\right)\cdot\boldsymbol{\sigma}^2\right]$$
(4.33)

These are written in terms of the body frame. However standard methods for the production of Bell states, using parametric down conversion for photons[17], the laboratory frame is defined by the crystal axes which filters out the antisymmetric terms, and leaves the invariant terms intact, *cf.* Eqs.(2.12) to (2.15). The relative orientation of the spins in each case is consistent with the angular momentum properties of the singlet and the triplet.

If no preferential laboratory axis exists, then over a random distribution the Bell states reduce to[5]:

$$\overline{\rho_{\psi_{12}^-}\left(\hat{\mathbf{n}}_{n_z,n_x}\right)}^{EA} = \frac{1}{4}\left(I^1 \otimes I^2 - \boldsymbol{\sigma}^1 \cdot \boldsymbol{\sigma}^2\right)$$
(4.34)

$$\overline{\rho_{\psi_{12}^+}\left(\hat{\mathbf{n}}_{n_z,n_x}\right)}^{EA} = \overline{\rho_{\Phi_{12}^-}\left(\hat{\mathbf{n}}_{n_z,n_x}\right)}^{EA} = \overline{\rho_{\Phi_{12}^+}\left(\hat{\mathbf{n}}_{n_z,n_x}\right)}^{EA} = \frac{1}{4}\left(I^1 \otimes I^2 + \frac{1}{3}\boldsymbol{\sigma}^1 \cdot \boldsymbol{\sigma}^2\right)$$
(4.35)

## 5. DISCUSSION

The non-hermitian states exist within hermitian bi-pariticle combinations Eqs.(2.7), (2.9) to (2.11), or in the presence of local and external interactions, whence only one axis of quantization survives, in agreement with the usual spin states obtained from quantum theory, Eq.(3.1). The appropriate sum of these bi-particle states leads to the hermitian Bell states which in quantum theory are entangled, but separate using the non-hermitian states. In contrast, isolated spins are predicted to have a resonance state, Eq.(3.5) which leads to hermitian spin states of magnitude $1/\sqrt{2}$. Although it is proposed below that some anomalous experiments are sensitive to coherent spins, in the majority of cases they have random phases and decoherence[14] renders the states hermitian. In short, the usual hermitian states of quantum mechanics are found to emerge when decoherence and ensemble averaging is warranted.

Quantum mechanics is a statistical theory[18,19,20] which means here that the states of individual particles can form ensembles which decohere and average. These individual and isolated spins are described by Eq.(3.5) and are well defined in the sense of EPR[9]. It is likely impossible, however, to distinguish these due to degeneracy found in the resonance states. These give a basis for understanding the statistical nature of quantum mechanics. Each degenerate state corresponds to a specific and well defined[9] orientation but for statistical filters, the degeneracy gives rationale for the inability to predict the outcomes of single events.

The emergence of the $\sqrt{2}$ spin rationalizes a number of experiments that quantum mechanics to date has not resolved. A common feature of these experiments is that the spins are isolated and in a magnetically isotropic environment.

Double slit experiments[21] show a build up of interference even when particles are fired ballistically. Since the non-hermitian operators have non-orthogonal eigenstates, Eq.(3.3), the treatment here gives a basis for an individual spin interfering with itself.

EPR coincidence experiments on photons[11,12] show violation of Bell's Inequalities[7] which is interpreted as supporting the notion of non-locality[8] These experiments are based on preparing spin in an entangled state and assuming the entanglement persists to space-like separations[22]. However within the constraints of quantum theory how the two particles conserve angular momentum after they are far apart, defies rational interpretation[23,24]. The use of non-hermitian spin states suggests that differently oriented bi-particles are produced and the statistical accumulation of these events is indistinguishable from the bound singlet, Eq.(4.15). This is due to the presence of a rotationally invariant contribution to all the Bell states, $\boldsymbol{\sigma}^1 \cdot \boldsymbol{\sigma}^2$. In particular every bi-particle formed by the decomposition of a Bell state retains only permutations of the invariant term in[25], Eqs.(4.7) and (4.31) to (4.33).

The original motivation for devising coincidence photon experiments[26] was to test Bell's Inequalities[7]. The existence of correlated bi-particles, in particular the presence of the √2 state, predicts that certain filter settings give maximum violations of Bell's Inequalities in coincidence experiments on photons[11,26,27] (*e.g.* 45° in the CHSH[28] form and 60° in Bell's original form[7] for spins of ½). In particular setting the filters at Alice and Bob at 45° shows strong correlation. This can be understood by the fact that the two separated spins, as they approach the filters at Alice and Bob share the same common axis in their body frame: $\boldsymbol{\sigma}^1 \cdot \hat{\mathbf{n}}$ and $\boldsymbol{\sigma}^2 \cdot \hat{\mathbf{n}}^*$, see Eqs.(4.10) and (4.11) Suppose Alice's filter is set at $\hat{\mathbf{a}}$, then in the laboratory frame the component along $\hat{\mathbf{n}}$ is

$$\hat{\mathbf{a}} \cdot \hat{\mathbf{n}} = \frac{1}{\sqrt{3}}\left(a_z + a_x + ia_y\right) = \frac{1}{\sqrt{3}} \begin{pmatrix} \cos\theta a_Z + \sin\theta\cos\varphi a_X + \sin\theta\sin\varphi a_Y \\ -\sin\theta a_Z + \cos\theta\cos\varphi a_X + \cos\theta\sin\varphi a_Y \\ -i\sin\varphi a_X + i\cos\varphi a_Y \end{pmatrix} \qquad (4.36)$$

Within the beam of particles a sub-ensemble can exist such that the laboratory and body frames share the same *y* axes, $\hat{\mathbf{y}} = \hat{\mathbf{Y}}$ so that $\varphi = 0$. For this sub-ensemble, the co-plane formed from the two filters coincides with the body frame plane formed from, $\hat{\mathbf{z}}$ and $\hat{\mathbf{x}}$ leading to

$$\left(\hat{\mathbf{a}} \cdot \hat{\mathbf{n}}\right)_{\phi=0} = \frac{1}{\sqrt{3}}\left(\left(\cos\theta - \sin\theta\right)a_Z + \left(\cos\theta + \sin\theta\right)a_X + ia_Y\right) \qquad (4.37)$$

Strong correlation is therefore predicted for any filter angles that differ by 90°. Since Bob's spin carries the same body frame, a setting at Alice and Bob differing by 90° will show strong correlations in coincidence counting.

There is, in addition, the √2 spin as seen in Eqs.(4.10) and (4.11). This bisects the spins body axes $\hat{\mathbf{z}}$ and $\hat{\mathbf{x}}$ and is filtered at the angle 45° between the two. Hence strong correlations are expected at laboratory filter settings that differ by 45° between Alice and Bob's filter settings, consistent with observations[22].

There are other filter settings that can be similarly predicted, in particular those of 60° and those of 180° (which suppress the √2 spin and exactly satisfy Bell's Inequalities.[22]) The 60° settings correspond to a sub-ensemble with the body frame axis, $\hat{\mathbf{n}}_{n_z n_x n_y}$, Eq.(4.5) perpendicular to the measurement co-plane. Since this axis trisects the body frame axes, they project onto the co-plane at 120°. However the $i\sigma_y$ component is not angular momentum and cannot be filtered while the projected $\hat{\mathbf{z}}$ and $\hat{\mathbf{x}}$ components can and show strong correlation if Alice and Bob's settings are 120° apart. Once again, the √2 spin bisects the projections of $\hat{\mathbf{z}}$ and $\hat{\mathbf{x}}$ onto the measurement co-plane leading to strong correlation when the filters are set at 60° apart. This analysis is not applicable for magnetic interactions, such as with a Stern-Gerlach filter, which destroys, rather than filters, the √2 spin.

Note that the imaginary part is not an observable; orients the body frame in 3D space; and becomes real in coincidence measurements, Eq.(4.13) which combines correlations between Alice and Bob after filtering.

The observed violations[11,12] have been viewed since Bell's work[8] as definitive evidence that any theories underpinning quantum mechanics must involve non-local connectivity. The treatment here puts into question that interpretation. In addition to his assumption of Einstein locality[9], Bell also assumed his classical events have values of only ±1. In contrast, as found here, isolated spins have values of ±√2, and this resonance and purely quantum spin state is responsible for the violations[27]. An alternative interpretation to non-locality is violations of Bell's inequalities is due to the existence of the √2 spin state, rather than violating the locality assumption.

In support of this, it has been shown mathematically by Gustafson[29] using non-commutative geometry, that violation of the CHSH[26] form of Bell's Inequalities is indeed due to a vector of length √2. The properties he finds for this vector agree with the predicted properties of the resonance spin state found here[27].

Although more work is needed, it is intriguing that non-hermitian spin states provide a logical and alternate framework for investigating these experiments. Quantum mechanics has been studied from many different formulations[30] and although these provide new insight, they do not provide new predictions, whereas the extension to non-hermitian states does. Moreover the non-hermiticity has clear physical meaning. The usual intrinsic spin angular momentum gives way to a two dimensional structure with degenerate resonance states that cancel in the absence of interactions. The non-hermiticity is essential for orienting a 2D spin in 3D real space, making each particle well defined[9]. These non-hermitian states are proposed to be the building blocks of angular momentum and their ensembles, and suggest the hermitian postulate be relaxed.

## ACKNOWLEDGMENT.

This work is supported by a Discovery Grant from the Natural Sciences and Engineering Research Council of Canada (NSERC).

## REFERENCES

[1] Edmonds, A. R., *Angular Momentum in Quantum Mechanics*., Princeton University Press, 1960.

[2] U. Fano, , "*Description of States of Quantum Mechanics by Density Matrix and Operator Techniques*." Rev. Mod. Phys. 1957 29, 74.

[3] Fano, U., *Pairs of two-level systems*. Rev. Mod. Phys. 1983, 55, 855.

[4] E. Schrödinger, *Discussion of Probability Relations Between Separated Systems*. Proceedings of the Cambridge Philosophical Society, 1935,31: 555-563.

[5] Many of the equations in this paper are derived, see http://www.mchmultimedia.com and look for "Equations for Separation of Bell States" at the bottom.

[6] See http://phet.colorado.edu/simulations/sims.php?sim=SternGerlach_Experiment

[7] J. S. Bell, *On the Einstein-Podolsky-Rosen Paradox*. Physics 1964 1 195.

[8] J. S. Bell, *Speakable and Unspeakable in Quantum Mechanics*, Cambridge University Press; New York, Cambridge University Press, 1987.

[9] A. Einstein, B. Podolsky, and N. Rosen, "Can quantum-mechanical description of physical reality be considered complete?" Phys. Rev. 1935 47, 777.

[10] E. Prugovečki, *Quantum Mechanics in Hilbert Space*, 1971 Academic Press (New York and London).

[11] G. Weihs, T. Jennewein, C. Simon, H. Weinfurter, A. Zeilinger, , "Violation of Bell's inequality under strict Einstein locality conditions", Phys.Rev.Lett. 1998, 81 5039-5043.

[12] A. Aspect, *Bell's Theorem: the Naïve View of an Experimentalist* in *Quantum [Un]speakables – From Bell to Quantum information*., edited by R. A. Bertlmann and A. Zeilinger, Springer 2002, 1–34.

[13] Adenier, G. and Khrennikov, A. Yu. *Is the fair sampling assumption supported by EPR experiments?*, J. Phrs.B: At. Mol. Opt. Phys. 40, 131-141 (2007).

[14] Ulrich Weiss, *Quantum Dissipative Systems*, in *Series in Modern Condensed Matter Physics-Vol. 13*. World Scientific, 3rd edition 2008.

[15] This section is derived, see reference 5.

[16] M. O. Scully, W. E. Lamb Jr., and A. Barut, *On the Theory of the Stern-Gerlach Apparatus*, Foundations of Physics, 1987, 17, 575.

[17] Kwiat, P. G., Mattle, K., Weinfurter, H. and Zeilinger, A., "New High-Intensity Source of Polarization-Entangled Photon Pairs", Phys. Rev. Lett., 1995, 75, 4337–4341.

[18] Jammer, Max, *The Philosophy of Quantum Mechanics*; *The interpretation of Quantum Mechanics in historical perspective*, John Wiley & son 1974.

[19] L. E. Ballentine, *The Statistical Interpretation of Quantum Mechanics*, *Rev. Mod. Phys*., 1970, 42, 358–381.

[20] L. E. Ballentine, *Quantum Mechanics, A Modern Development*. World Scientific Publishing Co. Ltd., 2000.

[21] Tonomura, A., Matsuda, J. Endo, T., Kawasaki, T. and Ezawa, H., *Demonstration of Single-Electron buildup of an Interference Pattern*, Am. J. Phys. 57, 117 (1989).

[22] J. F. Clauser and M. A. Horne, *Experimental consequences of objective local theories*. Phys. Rev. D, 1974, 10, 526-535.

[23] Private communication, N. Gisin, U. of Geneva, January 2005.

[24] Google quantum weirdness, quantum magic and spooky quantum

[25] $\psi_{12}^- : \hat{\mathbf{z}}^2 + \hat{\mathbf{x}}^2 + \hat{\mathbf{y}}^2$ ; $\psi_{12}^+ : -\hat{\mathbf{z}}^2 + \hat{\mathbf{x}}^2 + \hat{\mathbf{y}}^2$ ; $\phi_{12}^- : \hat{\mathbf{z}}^2 - \hat{\mathbf{x}}^2 + \hat{\mathbf{y}}^2$ ; $\phi_{12}^+ : \hat{\mathbf{z}}^2 + \hat{\mathbf{x}}^2 - \hat{\mathbf{y}}^2$

[26] J. F. Clauser and A. Shimony, *Bell's Theorem: Experimental Tests and Implications*, Rep. Prog. Phys., 1978, 41, 1881–1927.

[27] B. C. Sanctuary. *The two dimensional spin and its resonance fringe.* arXiv:0707.1763

[28] J. F. Clauser, M.A. Horne, A. Shimony and R. A. Holt, *Proposed experiment to test local hidden-variable theories*, Phys. Rev. Lett. 1969 23, 880-884.

[29] K.Gustafson, *Noncommutative Trigonometry and Quantum Mechanics*, in *Advances in Deterministic and Stochastic Analysis.* (N.Chuong, P.Ciarlet,P.Lax,D.Mumford, D.Phong, eds.), World Scientific 2007, p381.

[30] *Quantum Theory: reconsideration of foundations-3 International Conference*: June 6-11, 2005 International Center for Mathematical Modeling in Physics, Engineering and Cognitive Science, Växjö University, Sweden, A. Yu. Khrennikov organizer.